\documentclass[
    ,final            
  ]
  {aipproc}

\layoutstyle{8x11single}
\newcommand{\be}{\begin{equation}}
\newcommand{\ee}{\end{equation}}
\newcommand{\bea}{\begin{eqnarray}}
\newcommand{\eea}{\end{eqnarray}}
\newcommand{\bref}[1]{(\ref{#1})}

\begin{document}

\title{Renormalizable minimal SO(10) GUT in 4D and 5D\footnote{This is a talk in the Workshop on GUT held at Ritsumeikan Univ. on Dec.17-19 2007 \cite{Proceeding}}}


\author{Takeshi FUKUYAMA}{
  address={Department of Physics, Ritsumeikan University, Kusatsu, Shiga, 525-8577 Japan}
}



\begin{abstract}
This report is a review of the present status of GUT, especially renormalizable minimal SO(10) GUT, and its future prospect.
 It consists of two parts.
In part I, I review how the minimal renormalizable supersymmetric SO(10) GUT, 
 an SO(10) framework with only one ${\bf 10}$ and 
 one $\overline{\bf 126}$ Higgs multiplets 
 in the Yukawa sector, 
 is attractive because of its high predictivity.
Indeed it not only gave a consistent predictions on neutrino oscillation data but also did reasonable and interesting values for Leptogenesis,
LFV, muon g-2, neutrinoless double beta decay etc.
However, this model suffers from problems, apart from the small deviations from the observed values, related to running of gauge couplings and proton decay. 
The gauge coupling unification may be spoiled 
 due to the presence of intermediate scales much lighter than the grand unification (GUT) scale. 
In addition, the gauge couplings blow up around the GUT scale 
 because of the presence of Higgs multiplets of large representations. 
In order to remedy these pathologies, in part II, we extend GUT into 5D. We propose two approaches: one is to consider the warped extra dimension, using the bulk Higgs profile to explain the intermediate energy scales. 
Another is to use the orbifold GUT. Both approaches are complementary to each other.
\end{abstract}

\maketitle


\section{PART I --GUT in 4D}
SUSY GUT is the most promising candidate beyond the Standard Model (SM) \cite{GUT}.  
SM is the very powerful theory but it has the application limit like the other great theories.
There are discrepancies with observations like neutrino mass \cite{Murayama}, muon g-2,...
It also has the unsatisfactory points which strongly suggests more comprehensive theory: \\
SM has so many parameters. It does not explain quark-lepton mass spectra, mixing angles, all phases, Higgs mass stability against quantum correction, three different gauge couplings and their unification, and Dark Matters (DM) etc.\\
Here we consider the theory beyond SM from bottom up approach.
Top down approach from string theory is also interesting \cite{JEKim2}\cite{Kobayashi}.
On these problems there are many approaches but it seems to be SUSY GUT which
may solve the whole problems mentioned above. Of course there are still room to accept non SUSY GUT and even non GUT. \\
Even if we accept it, there are so many SUSY GUTs.
What is the gauge group, SU(5), SO(10), $E_6$ \cite{Maekawa}, $E_8\times E_8$ \cite{JEKim2} ?
So we need the other criterion to select the gauge symmetry.
Anomaly free condition may be good candidate for it.
Chiral symmetry must be preserved under quantum correction. SO(10)is the smallest group which is free from anomaly.
Such an anomaly free condition has well meaning if the theory is renormalizable.
Let us consider the structure of the exising theories. They have the form of
\be
L=L_{ren}+\frac{L_1}{\Lambda_1}.
\label{cutoff1}
\ee
Here the first $L_{ren}$ denotes renormalizable Lagrangian and the second unrenotmalizable effective Lagrangian. 
For $SU(3)\times U(1)$ theory, $L_1$ implies the Fermi coupling $G_WJ^\mu J_\mu$.
For $SU(3)\times SU(2)\times U(1)$ SM, this $L_1$ becomes the renormalizable
$g_2J^\mu W_\mu$ term but new effective term appears in the seesaw mechanism,
\be
L=L_{ren}'+Y_\nu^T\frac{1}{M_R}Y_\nu(LH)^2\equiv L_{ren}+\frac{L_2}{\Lambda_2}.
\label{cutoff2}
\ee
Here $L_{ren}'\supset L_{ren}'$ and $\Lambda_2(=O(10^{13}GeV))\gg \Lambda_1$. 
Thus the theory is expressed as the sum of renormalizable theory plus cut off effective action, and renormalizable Lagrangian becomes more involved as the energy scale goes high. In the limit of $\Lambda_2=\infty$ SM is renormalzable.
Thus it may be reasonable to consider the renormalizability as the guiding principle for model building beyond SM.

The group theoretical properties give strong constraints on quark-lepton of the same family but very weak on those of different families.
As for family symmetry, see \cite{King}.

The successful gauge coupling unification 
 in the minimal supersymmetric standard model (MSSM), 
strongly support the emergence of a supersymmetric (SUSY) GUT 
 around $M_{\rm GUT} \simeq 2 \times 10^{16}$ GeV. 
SO(10) is the smallest simple gauge group 
 under which the entire SM matter content of each generation 
 is unified into a single anomaly-free irreducible representation, 
${\bf 16}$ representation. 
This ${\bf 16}$ representation includes 
 right-handed neutrino and SO(10) GUT incorporates the see-saw 
mechanism \cite{see-saw}.  
 Among several models based on the gauge group SO(10), 
 the renormalizable minimal SO(10) model has been paid 
 a particular attention, where two Higgs multiplets 
 $\{{\bf 10} \oplus {\bf \overline{126}}\}$ 
 are utilized for the Yukawa couplings with matters 
 ${\bf 16}_i~(i=\mbox{generation})$.
A remarkable feature of the model is its high predictivity 
 of the neutrino oscillation parameters 
 as well as reproducing charged fermion masses and mixing angles.

\subsection{Minimal supersymmetric SO(10) model} 

First we give a brief review of the renormalizable minimal SUSY SO(10) model.\footnote{This part is based on the works by T. Fukuyama, A. Ilakovac, T. Kikuchi, S. Meljanac and N. Okada} 
\footnote{There is another flow of 
 non-renormalizable minimal SO(10) GUT  \cite{Babu}}.
\paragraph{Yukawa coupling}
This model was first applied to neutrino oscillation in \cite{Babu:1992ia}. 
However it did not reproduce the large mixing angles.
It has been pointed out that 
 CP-phases in the Yukawa sector play an important role 
 to reproduce the neutrino oscillation data \cite{Matsuda:2000zp}. 
More detailed analysis incorporating the renormalization group (RG) 
 effects in the context of MSSM 
 has explicitly shown that the model is consistent with the neutrino
 oscillation data at that time and became a realistic model \cite{Fukuyama:2002ch}. 
We give a brief review of this renormalizable minimal SUSY SO(10) model.\\
Yukawa coupling is given by
\begin{eqnarray}
 W_Y = Y_{10}^{ij} {\bf 16}_i H_{10} {\bf 16}_j 
           +Y_{126}^{ij} {\bf 16}_i H_{126} {\bf 16}_j \; , 
\label{Yukawa1}
\end{eqnarray} 
where ${\bf 16}_i$ is the matter multiplet of the $i$-th generation,  
 $H_{10}$ and $H_{126}$ are the Higgs multiplet 
 of {\bf 10} and $\overline{\bf 126} $ representations 
 under SO(10), respectively. 
Note that, by virtue of the gauge symmetry, 
 the Yukawa couplings, $Y_{10}$ and $Y_{126}$, 
 are, in general, complex symmetric $3 \times 3$ matrices. 
After the symmetry breaking pattern of SO(10) to 
${\rm SU}(3)_c \times {\rm SU}(2)_L \times {\rm U}(1)_Y$
via ${\rm SU}(4)_c \times {\rm SU}(2)_L \times {\rm SU}(2)_R$
or ${\rm SU}(5) \times {\rm U}(1)$,
 we find that two pair of Higgs doublets 
 in the same representation appear as the pair in the MSSM. 
One pair comes from $({\bf 1},{\bf 2},{\bf 2}) \subset {\bf 10}$ 
 and the other comes from 
 $(\overline{\bf 15}, {\bf 2}, {\bf 2}) \subset \overline{\bf 126}$. 
Using these two pairs of the Higgs doublets, 
 the Yukawa couplings of Eq.~(\ref{Yukawa1}) are rewritten as 
\begin{eqnarray}
W_Y &=& (U^c)_i  \left(
Y_{10}^{ij}  H^u_{10} + Y_{126}^{ij}  H^u_{126} \right) Q
+ (D^c)_i  \left(
Y_{10}^{ij}  H^d_{10} + Y_{126}^{ij}  H^d_{126} \right) Q_j  
\nonumber \\ 
&+& (N^c)_i \left( 
Y_{10}^{ij}  H^u_{10} - 3 Y_{126}^{ij} H^u_{126} \right) L_j 
+ (E^c)_i  \left(
Y_{10}^{ij}  H^d_{10}  - 3 Y_{126}^{ij} H^d_{126} \right) L_j   
\nonumber \\
&+&
 L_i \left( Y_{126}^{ij} \; v_T \right) L_j +
(N^c)_i \left( Y_{126}^{ij} \; v_R \right) (N^c)_j \;  , 
\label{Yukawa2}
\
\end{eqnarray} 
where $u_R$, $d_R$, $\nu_R$ and 
 $e_R$ are the right-handed ${\rm SU}(2)_L$ 
 singlet quark and lepton superfields, $q$ and $\ell$ 
 are the left-handed ${\rm SU}(2)_L$ doublet quark and lepton superfields, 
 $H_{10}^{u,d}$ and $H_{126}^{u,d}$ 
 are up-type and down-type Higgs doublet superfields 
 originated from $H_{10}$ and $H_{126}$, respectively, 
 and the last term is the Majorana mass term 
 of the right-handed neutrinos developed 
 by the VEV of the $(\overline{\bf 10}, {\bf 1}, {\bf 3})$ Higgs, $v_R$. 
The factor $-3$ in the lepton sector 
 is the Clebsch-Gordan coefficient. 

In order to preserve the successful gauge coupling unification, 
 suppose that one pair of Higgs doublets 
 given by a linear combination $H_{10}^{u,d}$ and $H_{126}^{u,d}$ 
 is light while the other pair is  heavy ($\geq M_{\rm GUT}$).  
The light Higgs doublets are identified as 
 the MSSM Higgs doublets ($H_u$ and $H_d$) 
 and given by 
\begin{eqnarray} 
 H_u &=& \tilde{\alpha}_u  H_{10}^u  
      + \tilde{\beta}_u  H_{126}^u \;,
 \nonumber \\
 H_d &=& \tilde{\alpha}_d  H_{10}^d  
      + \tilde{\beta}_d  H_{126}^d  \; , 
 \label{mix}
\end{eqnarray} 
where $\tilde{\alpha}_{u,d}$ and $\tilde{\beta}_{u,d}$ 
 denote elements of the unitary matrix  
 which rotate the flavor basis in the original model 
 into the (SUSY) mass eigenstates (See \bref{UV} in detail). 
Omitting the heavy Higgs mass eigenstates, 
 the low energy superpotential is described 
 by only the light Higgs doublets $H_u$ and $H_d$ such that 
\begin{eqnarray}
W_Y &=& 
(U^c) _i \left( \alpha^u  Y_{10}^{ij} + 
\beta^u Y_{126}^{ij} \right)  H_u \, Q_j 
+ (D^c)_i  
\left( \alpha^d  Y_{10}^{ij} + 
\beta^d Y_{126}^{ij}  \right) H_d \,Q_j  \nonumber \\ 
&+& (N^c)_i  
\left( \alpha^u  Y_{10}^{ij} -3 
\beta^u Y_{126}^{ij} \right)  H_u \,L_j 
+ (E^c)_i  
\left( \alpha^d  Y_{10}^{ij} -3 
\beta^d  Y_{126}^{ij}  \right) H_d \,L_j \nonumber \\ 
&+& 
  L_i \left( Y_{126}^{ij} \; v_T \right) L_j + 
 (N^c)_i  
  \left( Y_{126}^{ij} v_R \right)  (N^c)_j \; ,  
\label{Yukawa3}
\end{eqnarray}
where the formulas of the inverse unitary transformation 
 of Eq.~(\ref{mix}), 
 $H_{10}^{u,d} = \alpha^{u,d} H_{u,d} + \cdots $ and 
 $H_{126}^{u,d} = \beta^{u,d} H_{u,d} + \cdots $, 
 have been used. 
Note that the elements of the unitary matrix, 
 $\alpha^{u,d}$ and $\beta^{u,d}$,   
 are in general complex parameters, 
 through which CP-violating phases are introduced 
 into the fermion mass matrices. 

Providing the Higgs VEVs, 
 $H_u = v \sin \beta$ and $H_d = v \cos \beta$ 
 with $v=174 \mbox{GeV}$, 
 the quark and lepton mass matrices can be read off as%
\begin{eqnarray}
  M_u &=& c_{10} M_{10} + c_{126} M_{126}   \nonumber \\
  M_d &=&     M_{10} +     M_{126}   \nonumber \\
  M_D &=& c_{10} M_{10} -3 c_{126} M_{126}   \nonumber \\
  M_e &=&     M_{10} -3     M_{126}   \nonumber \\
  M_T &=& c_T M_{126} \nonumber \\  
  M_R &=& c_R M_{126}  \; , 
 \label{massmatrix}
\end{eqnarray} 
where $M_u$, $M_d$, $M_D$, $M_e$, $M_T$, and $M_R$ 
 denote the up-type quark, down-type quark, 
 Dirac neutrino, charged-lepton, left-handed Majorana, and 
 right-handed Majorana neutrino mass matrices, respectively. 
Note that all the quark and lepton mass matrices 
 are characterized by only two basic mass matrices, $M_{10}$ and $M_{126}$,   
 and four complex coefficients 
 $c_{10}$, $c_{126}$, $c_T$ and $c_R$, 
 which are defined as 
 $M_{10}= Y_{10} \alpha^d v \cos\beta$, 
 $M_{126} = Y_{126} \beta^d v \cos\beta$, 
 $c_{10}= (\alpha^u/\alpha^d) \tan \beta$, 
 $c_{126}= (\beta^u/\beta^d) \tan \beta $, 
 $c_T = v_T/( \beta^d  v  \cos \beta)$) and 
 $c_R = v_R/( \beta^d  v  \cos \beta)$), respectively.  
These are the mass matrix relations required by 
 the minimal SO(10) model. 
In the following in Part I, we set $c_T=0$ as the first approximation.
Except for $c_R$, 
  which is used to determine the overall neutrino mass scale, 
 this system has fourteen free parameters in total \cite{Matsuda:2000zp}, 
 and the strong predictability to the fermion mass matrices.
The reasonable results we found are listed in Table 1.
\begin{table}[pt]
\caption{The input values of $\tan \beta$, $m_s(M_Z)$
and $\delta$ in the CKM matrix and the outputs for the neutrino
oscillation parameters.}
{\begin{tabular}{c|cc|c|ccc|c}
\hline \hline
 $\tan \beta $ & $m_s(M_Z)$ & $\delta$  & $\sigma $
 & $\sin^2 2 \theta_{1 2}$
 & $\sin^2 2 \theta_{2 3}$
 & $\sin^2 2 \theta_{1 3} $
 & $\Delta m_{\odot}^2/\Delta m_{\oplus}^2$ \\ \hline
40 & 0.0718 & $ 93.6^\circ $ & 3.190& 
0.738 & 0.900 & 0.163 & 0.205 \\
45 & 0.0729 & $ 86.4^\circ $ & 3.198& 
0.723 & 0.895 & 0.164 & 0.188 \\
50 & 0.0747 & $ 77.4^\circ $ & 3.200& 
0.683 & 0.901 & 0.164 & 0.200 \\
55 & 0.0800 & $ 57.6^\circ $ & 3.201& 
0.638 & 0.878 & 0.152 & 0.198 \\
\hline \hline
\end{tabular}}
\end{table}
\begin{figure}
  \includegraphics[height=.3\textheight]{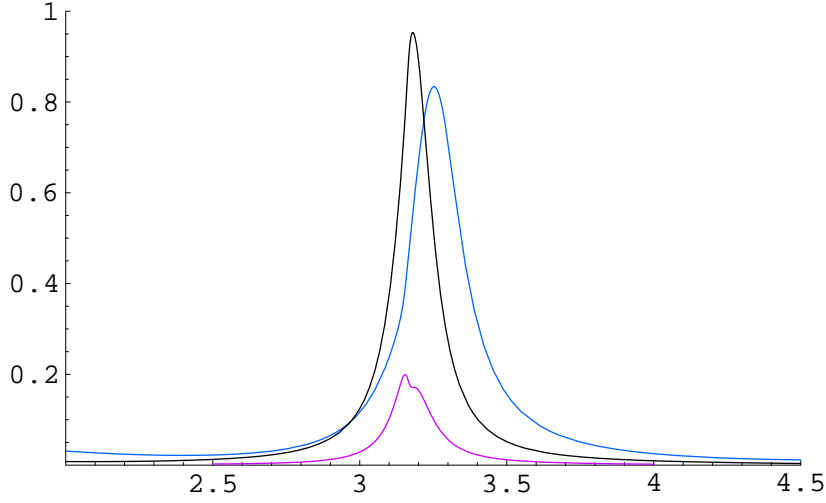}
  \caption{Three mixing angles in the MNS matrix as functions of $\sigma \mbox{[rad]}$. 
The graphs with the highest, middle and lowest peaks
correspond to $\sin^2 2 \theta_{2 3}$, 
$\sin^2 2 \theta_{1 2}$ and $\sin^2 2 \theta_{1 3}$, respectively. 
The plots of $\sin^2 2 \theta_{23}$ and $\sin^2 2 \theta_{13}$ have 
the sharp peaks at $\sigma \sim 3.2 [\mbox{rad}]$, 
while $\sin^2 2 \theta_{12}$ has the sharp peak at 
$\sigma \sim 3.3 [\mbox{rad}]$ cited from \cite{Fukuyama:2002ch}}
\end{figure}

As mentioned above, 
 our resultant neutrino oscillation parameters 
 are sensitive to all the input parameters. 
In other words, if we use the neutrino oscillation data 
 as the input parameters, 
 the other input, for example, the CP-phase in the CKM matrix 
 can be regarded as the prediction of our model. 
It is a very interesting observation 
 that the CP-phases listed above are 
 in the region consistent with experiments.
The CP-violation in the lepton sector
 is characterized by the Jarlskog parameter
 defined as 
\begin{eqnarray}
 J_{CP} = \mbox{Im}\left[ 
     U_{e2} U_{\mu 2}^*  U_{e 3}^*  U_{\mu 3} \right] \; , 
\end{eqnarray}  
where $U_{f i}$ is the MNS matrix element. 

\paragraph{Lepton Flavour Violation}
It is well known that the SO(10) GUT model possesses 
 a simple mechanism of baryogenesis 
 through the out-of-equilibrium decay 
 of the right-handed neutrinos, 
 namely, the leptogenesis \cite{Fukugita-Yanagida}. 
The lepton asymmetry in the universe is generated by CP-violating 
out-of-equilibrium decay of the heavy neutrinos,  
$N \rightarrow \ell_L H_u^*$ and $N \rightarrow \overline{\ell_L} H_u$. 
The leading contribution is given by the interference between 
the tree level and one-loop level decay amplitudes, 
and the CP-violating parameter is found to be
\begin{eqnarray}
\epsilon = 
\frac{1}{8 \pi (Y_\nu Y_\nu^\dag)_{11}}
\sum_{j=2,3}\mbox{Im} \left[ (Y_\nu Y_\nu^\dag)_{1j}^2 \right]
\left\{ f(M_{Rj}^2/M_{R1}^2)
+ 2 g(M_{Rj}^2/M_{R1}^2) \right\} \; .
\label{epsilon}
\end{eqnarray}
Here $f(x)$ and $g(x)$ correspond to 
the vertex and the wave function corrections, 
\begin{eqnarray}
f(x)&\equiv& \sqrt{x} \left[
1-(1+x)\mbox{ln} \left(\frac{1+x}{x} \right) \right] \;,
\nonumber\\
g(x)&\equiv& \frac{\sqrt{x}}{2(1-x)}   \; ,  
\end{eqnarray}
respectively, and both are reduced to 
$\sim -\frac{1}{2 \sqrt{x}}$ for $ x \gg 1$. 
So in this approximation, $\epsilon$ becomes 
\begin{equation}
\epsilon = - 
\frac{3}{16 \pi (Y_\nu Y_\nu^\dag)_{11}}
\sum_{j=2,3} \mbox{Im} \left[(Y_\nu Y_\nu^\dag)_{1j}^2 \right]
\frac{M_{R1}}{M_{Rj}}\;.
\end{equation}
These quantities are evaluated by using the results 
 presented in Table 1, 
 and the results are listed in Table 2.  
\begin{table}[pt]
\caption{The input values of $\tan \beta$ and the outputs 
for the CP-violating observables}
{\begin{tabular}{c|ccc}
\hline \hline
 $\tan \beta $ & 
 $ \langle m_\nu \rangle_{ee}~ (\mbox{eV})$ & 
 $J_{CP}$ &   $\epsilon$  \\   \hline
40 & 0.00122  & $~~0.00110$ & $ 7.39 \times 10^{-5} $ \\
45 & 0.00118  & $-0.00429$  & $ 6.80 \times 10^{-5} $ \\
50 & 0.00119  & $-0.00631$  & $ 6.50 \times 10^{-5} $ \\
55 & 0.00117  & $-0.00612$  & $ 11.2 \times 10^{-5} $ \\ 
\hline \hline
\end{tabular}}
\end{table}

Now we turn to the discussion about the rate of the LFV processes
and the muon $g-2$.
The evidence of the neutrino flavor mixing implies that 
 the lepton flavor of each generation is not individually conserved. 
Therefore the lepton flavor violating (LFV) processes 
 in the charged-lepton sector such as 
 $\mu \rightarrow e \gamma$, $\tau \rightarrow \mu \gamma$ 
 are allowed. 
In simply extended models 
 so as to incorporate massive neutrinos into the standard model, 
 the rate of the LFV processes is accompanied 
 by a highly suppression factor, 
 the ratio of neutrino mass to the weak boson mass, 
 because of the GIM mechanism,  
 and is far out of the reach of the experimental detection. 
However, in supersymmetric models, the situation is quite different. 
In this case, soft SUSY breaking parameters can be new LFV sources, 
 and the rate of the LFV processes 
 are suppressed by only the scale 
 of the soft SUSY breaking parameters 
 which is assumed to be the electroweak scale. 
Thus the huge enhancement occurs compared to the previous case. 
In fact, the LFV processes can be one of the most important processes 
 as the low-energy SUSY search. 
we evaluate the rate of the LFV processes 
 in the minimal SUSY SO(10) model, 
 where the neutrino Dirac Yukawa couplings 
 are the primary LFV sources. 
Although in Ref.~\cite{Fukuyama:2002ch} 
 various cases with given $\tan \beta = 40-55$ have been analyzed, 
 we consider only the case $\tan \beta =45$ in the following. 
Our final result is almost insensitive 
 to $\tan \beta$ values in the above range. 
The predictions of the minimal SUSY SO(10) model 
 necessary for the LFV processes are as follows \cite{Fukuyama:2002ch}: 
 with $\sigma=3.198$ fixed, 
 the right-handed Majorana neutrino mass eigenvalues 
 are found to be (in GeV) 
 $M_{R_1}=1.64 \times 10^{11}$,  
 $M_{R_2}=2.50 \times 10^{12}$ and 
 $M_{R_3}=8.22 \times 10^{12}$, 
 where $c_R$ is fixed so that 
 $\Delta m_\oplus^2 = 2 \times 10^{-3} \mbox{eV}^2$. 
In the basis where both of the charged-lepton 
 and right-handed Majorana neutrino mass matrices 
 are diagonal with real and positive eigenvalues, 
 the neutrino Dirac Yukawa coupling matrix at the GUT scale 
 is found to be \footnote{We are now reconsidering data fitting with
the up todate experimental data and new RGE results. It gives the differen values from \bref{Ynu} but the LFV results are not essentially changed.}
\begin{eqnarray}
 Y_{\nu} = 
\left( 
 \begin{array}{ccc}
-0.000135 - 0.00273 i & 0.00113  + 0.0136 i  & 0.0339   + 0.0580 i  \\ 
 0.00759  + 0.0119 i  & -0.0270   - 0.00419  i  & -0.272    - 0.175   i  \\ 
-0.0280   + 0.00397 i & 0.0635   - 0.0119 i  &  0.491  - 0.526 i 
 \end{array}   \right) \; .  
\label{Ynu}
\end{eqnarray}     
LFV effect most directly emerges 
 in the left-handed slepton mass matrix 
 through the RGEs such as \cite{Hisano}
\begin{eqnarray}
\mu \frac{d}{d \mu} 
  \left( m^2_{\tilde{\ell}} \right)_{ij}
&=&  \mu \frac{d}{d \mu} 
  \left( m^2_{\tilde{\ell}} \right)_{ij} \Big|_{\mbox{MSSM}} 
 \nonumber \\
&+& \frac{1}{16 \pi^2} 
\left( m^2_{\tilde{\ell}} Y_{\nu}^{\dagger} Y_{\nu}
 + Y_{\nu}^{\dagger} Y_{\nu} m^2_{\tilde{\ell}} 
 + 2  Y_{\nu}^{\dagger} m^2_{\tilde{\nu}} Y_{\nu}
 + 2 m_{H_u}^2 Y_{\nu}^{\dagger} Y_{\nu} 
 + 2  A_{\nu}^{\dagger} A_{\nu} \right)_{ij}  \; ,
 \label{RGE} 
\nonumber\\
\end{eqnarray}
where the first term in the right hand side denotes 
 the normal MSSM term with no LFV. 
We have found $Y_\nu$ explicitly and we can calculate LFV and
 related phenomena unambiguously \cite{fukuyama2}
In the leading-logarithmic approximation, 
 the off-diagonal components ($i \neq j$)
 of the left-handed slepton mass matrix are estimated as 
\begin{eqnarray}
 \left(\Delta  m^2_{\tilde{\ell}} \right)_{ij}
 \sim - \frac{3 m_0^2 + A_0^2}{8 \pi^2} 
 \left( Y_{\nu}^{\dagger} L Y_{\nu} \right)_{ij} \; ,  
 \label{leading}
\end{eqnarray}
where the distinct thresholds of the right-handed 
 Majorana neutrinos are taken into account 
 by the matrix $ L = \log [M_{\rm GUT}/M_{R_i}] \delta_{ij}$. 

The effective Lagrangian 
 relevant for the LFV processes ($\ell_i \rightarrow \ell_j \gamma$) 
 and the muon $g-2$ is described as 
\begin{eqnarray}
 {\cal L}_{\mbox{eff}}= 
 -  \frac{e}{2} m_{\ell_i} \overline{\ell}_j \sigma_{\mu \nu} F^{\mu \nu} 
 \left(A_L^{j i} P_L + A_R^{j i} P_R \right) \ell_i  \; , 
\end{eqnarray}
where $P_{R, L} = (1 \pm \gamma_5)/2 $ is  
 the chirality projection operator, 
 and  $A_{L,R}$ are the photon-penguin couplings of 1-loop diagrams 
 in which chargino-sneutrino and neutralino-charged slepton 
 are running. 
The explicit formulas of $A_{L,R}$ etc. used in our analysis 
 are summarized in \cite{Hisano-etal} \cite{Okada-etal}. 
The rate of the LFV decay of charged-leptons is given by 
\begin{eqnarray}
\Gamma (\ell_i \rightarrow \ell_j \gamma) 
= \frac{e^2}{16 \pi} m_{\ell_i}^5 
 \left( |A_L^{j i}|^2  +  |A_R^{j i}|^2  \right) \; , 
\end{eqnarray}
while the real diagonal components of $A_{L,R}$ 
 contribute to the anomalous magnetic moments of 
 the charged-leptons such as 
\begin{eqnarray}
 \delta a_{\ell_i}^{\rm SUSY} = \frac{g_{\ell_i}-2}{2} 
  = -  m_{\ell_i}^2 
  \mbox{Re} \left[ A_L^{i i}  +  A_R^{i i}  \right]  \; . 
\end{eqnarray}
In order to clarify the parameter dependence 
 of the decay amplitude, 
 we give here an approximate formula of the LFV decay rate 
 \cite{Hisano-etal}, 
\begin{eqnarray}
\Gamma (\ell_i \rightarrow \ell_j \gamma) 
 \sim  \frac{e^2}{16 \pi} m_{\ell_i}^5 
 \times  \frac{\alpha_2}{16 \pi^2} 
 \frac{\left| \left(\Delta  m^2_{\tilde{\ell}} \right)_{ij}\right|^2}{M_S^8} 
 \tan^2 \beta \; , 
 \label{LFVrough}
\end{eqnarray}
where $M_S$ is the average slepton mass at the electroweak scale, 
 and $ \left(\Delta  m^2_{\tilde{\ell}} \right)_{ij}$ 
 is the slepton mass estimated in Eq.~(\ref{leading}). 
We can see that the neutrino Dirac Yukawa coupling matrix 
 plays the crucial role in calculations of the LFV processes. 
We use the neutrino Dirac Yukawa coupling matrix of Eq.~(\ref{Ynu})
 in our numerical calculations. 

\begin{figure}[th]
\includegraphics[width=5cm]{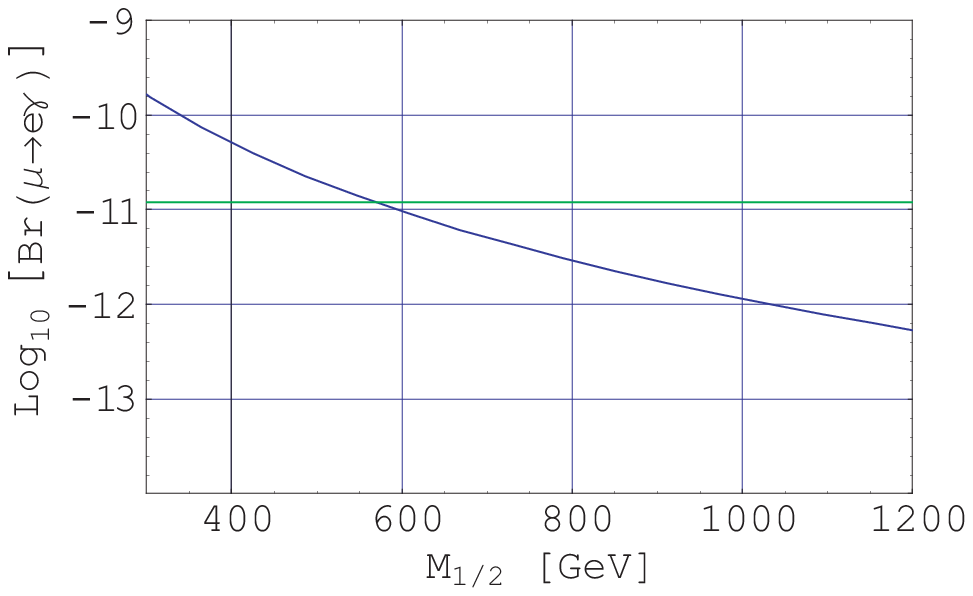}
\label{Fig2a}
\includegraphics[width=5cm]{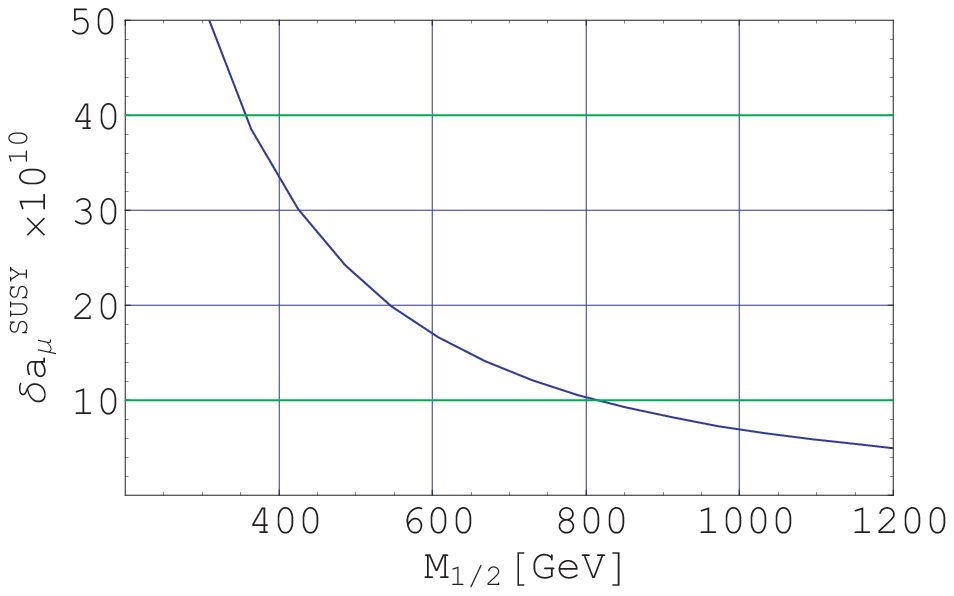}
\label{Fig2b}
\includegraphics[width=5cm]{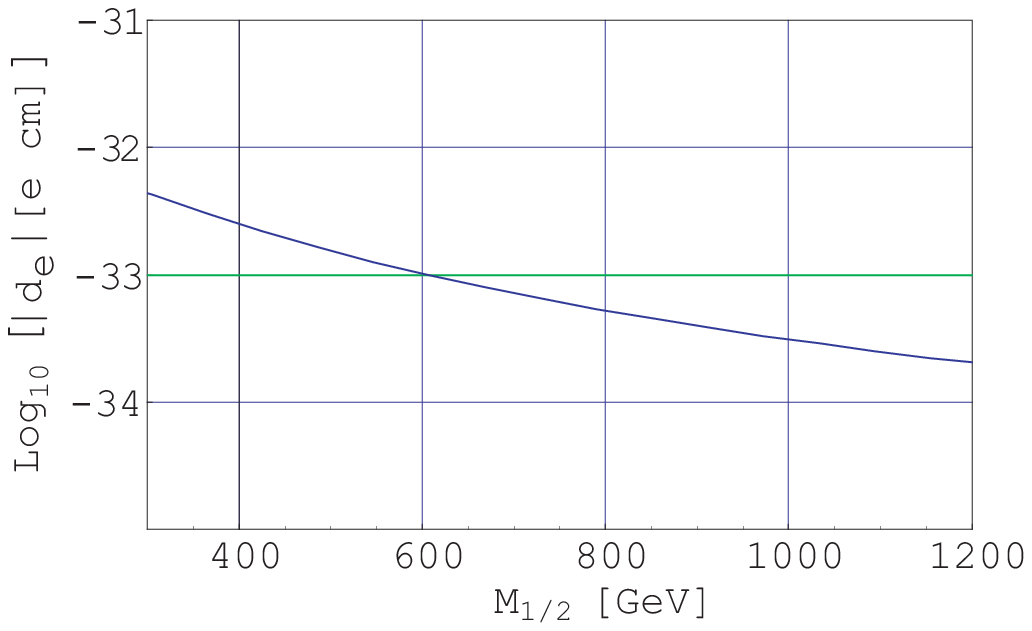}\label{Fig2c}
\vspace*{8pt}
\caption{
The branching ratio, a:
$\mbox{Log}_{10} \left[{\rm Br} (\mu \rightarrow e \gamma) \right]$, b:
the SUSY contribution to the muon $g-2$ in units of $10^{-10}$, 
$\delta a_{\ell_i}^{\rm SUSY} = \frac{g_{\ell_i}-2}{2}$,
and c:
the electron EDM, 
$\mbox{Log}_{10} \left[ | d_e | [ \mbox{e cm}] \right]$. 
All these figures are plotted as a function of $M_{1/2}$ (GeV) 
 along the cosmological constraint of Eq.~(\ref{relation}).
}
\end{figure}

The recent Wilkinson Microwave Anisotropy Probe (WMAP) satellite data 
 \cite{WMAP}  
 provide estimations of various cosmological parameters 
 with greater accuracy. 
The current density of the universe is composed of 
 about 73\% of dark energy and 27\% of matter. 
Most of the matter density is in the form of 
 the CDM, and its density is estimated to be (in 2$\sigma$ range) 
\begin{eqnarray}
\Omega_{\rm CDM} h^2 = 0.1126^{+0.0161}_{-0.0181} \; . 
 \label{WMAP} 
\end{eqnarray}
The parameter space of the CMSSM 
 which allows the neutralino relic density 
 suitable for the cold dark matter 
 has been recently re-examined 
 in the light of the WMAP data \cite{CDM}. 
It has been shown that the resultant parameter space 
 is dramatically reduced into the narrow stripe 
 due to the great accuracy of the WMAP data. 
It is interesting to combine this result 
 with our analysis of the LFV processes and the muon $g-2$. 
In the case relevant for our analysis, 
 $\tan \beta=45$, $\mu>0$ and $A_0=0$, 
 we can read off the approximate relation 
 between $m_0$ and $M_{1/2}$ 
 such as (see Figure 1 in the second paper of Ref.~\cite{CDM}) 
\begin{eqnarray}
m_0(\mbox{GeV}) = \frac{9}{28} M_{1/2}(\mbox{GeV}) + 150(\mbox{GeV}) \;,  
 \label{relation} 
\end{eqnarray} 
along which the neutralino CDM is realized. 
$M_{1/2}$ parameter space is constrained within the range 
 $300 \mbox{GeV} \leq M_{1/2} \leq 1000 \mbox{GeV}$ 
 due to the experimental bound on the SUSY contribution 
 to the $ b \rightarrow s \gamma$ branching ratio 
 and the unwanted stau LSP parameter region. 
We show $\mbox{Br}(\mu \rightarrow e \gamma)$ and 
 the muon $g-2$ as functions of $M_{1/2}$ 
 in Fig.~\ref{Fig2a} and \ref{Fig2b}, respectively, 
 along the neutralino CDM condition of Eq.~(\ref{relation}). 
We find the parameter region, 
 $560 \mbox{GeV} \leq M_{1/2} \leq 800 \mbox{GeV}$, 
 being consistent with all the experimental data. 

%
%

The semileptonic flavor violation processes were also considered
in \cite{fukuyama3}, for instance, for 
$\tau^- \to e^- (\mu^-)\, \pi^0$,
$\tau^- \to e^- (\mu^-)\, \eta$,
$\tau^- \to e^- (\mu^-)\, \eta^\prime$,
$\tau^- \to e^- (\mu^-)\, \rho^0$,
$\tau^- \to e^- (\mu^-)\, \phi$,
$\tau^- \to e^- (\mu^-)\, \omega$, {\it etc.}

When the KamLAND data \cite{Eguchi:2002dm} was released, 
 the results in \cite{Fukuyama:2002ch} were found to be deviated 
 by 3$\sigma$ from the observations. 
Afterward this minimal SO(10) was modified by many authors, 
 using the so-called type-II see-saw mechanism \cite{Bajc:2002iw} 
 and/or considering a ${\bf 120}$ Higgs coupling to the matter
 in addition to the ${\bf \overline{126}}$ Higgs \cite{Goh:2003hf}. 
Based on an elaborate input data scan \cite{Babu:2005ia}, 
\cite{Bertolini:2006pe},
 it has been shown that the minimal SO(10) is essentially consistent 
 with low energy data of fermion masses and mixing angles. 
The importance of the threshold
 corrections was also discussed in \cite{Parida}
\paragraph{Higgs Superpotential}
On the other hand, it has been long expected to construct 
 a concrete Higgs sector of the minimal SO(10) model. 
The simplest Higgs superpotential at the renormalizable level
is given by \cite{clark}, \cite{lee}, \cite{aulakh}
\be
W=m_1 \Phi^2 + m_2 \Delta \overline{\Delta} 
+m_3 H^2
+\lambda_1 \Phi^3 + \lambda_2 \Phi \Delta \overline{\Delta}
+\lambda_3 \Phi \Delta H + \lambda_4 \Phi \overline{\Delta} H \;,
\label{lee}
\ee
where $\Phi ={\bf 210}$, $\Delta ={\bf 126}$, 
$\overline{\Delta} ={\bf \overline{126}}$ and $H={\bf 10}$.
The interactions of ${\bf 210}$, ${\bf \overline{126}}$, 
${\bf 126}$ and ${\bf 10}$ lead to some complexities 
in decomposing the GUT representations to the MSSM 
and in getting the low energy mass spectra.  
Particularly, the CG coefficients 
corresponding to the decompositions of 
${\rm SO}(10) \to 
{\rm SU}(3)_C \times {\rm SU}(2)_L \times {\rm U}(1)_Y$ 
have to be found.
This problem was first attacked by X.~G.~He and S.~Meljanac
\cite{he} and further by D.~G.~Lee \cite{lee} and by J.~Sato \cite{Joe}. 
But they did not present the explicit form of mass matrices for 
a variety of Higgs fields and also did not perform a formulation 
of the proton life time analysis.  This is very labourious work and it is indispensable for the data fit of low energy physics.
We completed that program in \cite{Fukuyama:2004xs} (See also \cite{Bajc:2004xe},
\cite{Aulakh:2004hm}).
This construction gives some constraints among the vacuum expectation
 values (VEVs) of several Higgs multiplets, 
 which give rise to a trouble in the gauge coupling unification 
 \cite{Bertolini:2006pe}. 
The trouble comes from the fact that the observed neutrino oscillation
 data suggests the right-handed neutrino mass around $10^{13-14}$ GeV,
 which is far below the GUT scale. Indeed \bref{lee} contains 
five directions which are singlets 
under $SU(3)_C \times SU(2)_L \times U(1)_Y$.  
Three of them are included in {\bf 210}, 
\bea
\hat{\phi}_1 &=& (1234), 
\\
\hat{\phi}_2 &=& (5678+5690+7890), 
\\
\hat{\phi}_3 &=& (12+34)(56+78+90).  
\eea
one in {\bf 126} and 
\be
\hat{v}_R = (13579), 
\ee
and one in ${\bf \overline{126}}$
\be
\hat{\overline{v_R}} = (24680).  
\ee

Due to the D-flatness condition the VEVs 
$v_R$ and $\overline{v_R}$ are equal (\ref{lee}), 
\be
v_R= \overline{v_R}. 
\ee

This intermediate scale is provided by Higgs field VEV, 
 and several Higgs multiplets are expected to have their masses 
 around the intermediate scale and contribute to 
 the running of the gauge couplings. 

We write down the VEV conditions which preserve supersymmetry, 
with respect to the directions 
$\hat{\phi}_1$, $\hat{\phi}_2$, $\hat{\phi}_3$, and $\hat{v}_R$, 
respectively.  
\bea
\label{VEV1}
2 m_1 \phi_1 + 3 \lambda_1 \frac{ \phi_3^2}{6 \sqrt{6}} 
+ \lambda_2 \frac{v_R^2}{10 \sqrt{6}} &=& 0,
\\
\label{VEV2}
2 m_1 \phi_2 + 3 \lambda_1 \left(\frac{\phi_2^2 +\phi_3^2}
{9 \sqrt{2}} \right)+ \lambda_2 \frac{v_R^2}{10 \sqrt{2}} &=& 0,
\\
\label{VEV3}
2 m_1 \phi_3 + 3 \lambda_1 \left(\frac{\phi_1 \phi_3}{3 \sqrt{6}}
+\frac{\sqrt{2} \phi_2 \phi_3}{9}\right)
+ \lambda_2 \frac{v_R^2}{10} &=& 0,
\\
\label{VEV4}
m_2 + \lambda_2 \left( \frac{\phi_1}{10 \sqrt{6}}
+ \frac{\phi_2}{10 \sqrt{2}}
+ \frac{\phi_3}{10} \right)&=& 0.
\label{VEV}
\eea

Eliminating $v_R^2$, $\phi_1$ and $\phi_3$ from 
Eqs. (\ref{VEV1})--(\ref{VEV4}), 
one obtains a fourth-order equation in $\phi_2$.  
The corresponding fourth-order polynomial in $\phi_2$ 
factorizes into a linear and a cubic term 
in $\phi_2$.  
Linear term gives the solution of the fourth-order equation 
which is very simple, $\phi_2 = -3 \sqrt{2} \, m_2/\lambda_2$, 
but it preserves the $SU(5)$ symmetry.  
Therefore, it is physically not 
interesting.  The cubic term solutions lead to the true 
$SU(3)_C \times SU(2)_L \times U(1)_Y$ symmetry.  
Here we consider only the solutions with $|v_R| \neq 0$.
Eliminating $v_R \cdot \overline{v_R}$, 
$\phi_1$ and $\phi_2$ from 
Eqs. (\ref{VEV1})--(\ref{VEV4}), 
one obtains a fourth-order equation in $\phi_3$,  
\be
\label{phi3eq}
\left(\phi_3 + \frac{{\cal M}_2}{10} \right)
\left\{8 \, \phi_3^3 - 15 \, {\cal M}_1 \phi_3^2 
+ 14 \, {\cal M}_1^2 \phi_3 - 3 \, {\cal M}_1 ^3 
+ \left(\phi_3 - {\cal M}_1 \right)^2 {\cal M}_2 \right\} 
= 0,  
\ee
where 
\be
{\cal M}_1 \equiv 12 \, \left(\frac{m_1}{\lambda_1} \right), \, 
{\cal M}_2 \equiv 60 \, \left(\frac{m_2}{\lambda_2} \right).  
\ee
Any solution of the cubic equation in $\phi_3$ is 
accompanied by the solutions 
\bea
\phi_1 &=& - \frac{\phi_3}{\sqrt{6}} 
\frac{\left({\cal M}_1^2 - 5 \, \phi_3^2 \right)}{({\cal M}_1 - \phi_3)^2}, 
\\
\phi_2 &=& - \frac{1}{\sqrt{2}} 
\frac{\left({\cal M}_1^2 - 2 \, {\cal M}_1 \phi_3 - \phi_3^2 \right)}{({\cal M}_1 - \phi_3)}, 
\\
v_R \cdot \overline{v_R} &=&  \frac{5}{3} \, 
\left(\frac{\lambda_1}{\lambda_2} \right) 
\frac{\phi_3 \left({\cal M}_1 - 3 \, \phi_3 \right) \left({\cal M}_1^2 + \phi_3^2 \right)}
{({\cal M}_1 - \phi_3)^2}.  
\label{nuR}
\eea
The linear term gives the solution of the fourth-order equation (\ref{phi3eq})
which is very simple, 
$\phi_3 = - 6 \, \left(\frac{m_2}{\lambda_2} \right)$.
It leads to $\phi_1 = -\sqrt{6} \, \left(\frac{m_2}{\lambda_2} \right)$, 
$\phi_2 = -3 \sqrt{2} \, \left(\frac{m_2}{\lambda_2} \right)$    and 
$\sqrt{\left( v_R \cdot \overline{v_R} \right)} = \sqrt{60} \, 
\left(\frac{m_2}{\lambda_2} \right) \sqrt{2 \left(\frac{m_1}{m_2} \right) 
- 3 \left(\frac{\lambda_1}{\lambda_2} \right)}$.  
This solution preserves the $SU(5)$ symmetry.  
Therefore, it is physically not interesting.  
Then we proceed to the most important part of the SO(10) GUT.
We can not show the detail of the scenario but only show the essential part of it \cite{Fukuyama:2004xs}. 
\paragraph{Would-be Nambu-Goldstone bosons}

At first, we list the quantum numbers of 
the would-be NG modes under $SU(3)_C \times SU(2)_L \times U(1)_Y$.  

\begin{itemize}
\item
$
\left[
\left({\bf \overline{3}, 2},\frac{5}{6} \right)
\oplus
\left({\bf 3, 2}, -\frac{5}{6} \right)
\right],
$
\item 
$
\left[
\left({\bf \overline{3}, 2},-\frac{1}{6} \right)
\oplus
\left({\bf 3, 2},\frac{1}{6} \right)
\right],
$
\item
$
\left[
\left({\bf \overline{3}, 1},-\frac{2}{3} \right)
\oplus
\left({\bf 3, 1},\frac{2}{3} \right)
\right], 
$
\item
$
\left[
\left({\bf 1, 1},1 \right)
\oplus
\left({\bf 1, 1},-1 \right)
\right],
$
\item
$\left[\left({\bf 1, 1},0 \right)\right].$
\end{itemize}
Total number of the NG degrees of freedom is :  
${\bf 12 + 12 + 6 + 2 + 1 =  33}$. 
The cubic term solutions lead to the true 
$SU(3)_C \times SU(2)_L \times U(1)_Y$ symmetry.  
\bref{nuR} gives heavy right-handed neutrino, and the coefficient of \bref{massmatrix} is also written in terms of $\phi_3$.
\paragraph{Electroweak Higgs doublet}
In the standard picture of the electroweak symmetry breaking, 
we have the Higgs doublets which give masses to the matter.  
These masses should be less than or equal to the electroweak scale.  
Since we approximate the electroweak scale as zero, 
we must impose a constraint that the mass matrix should 
have one zero eigenvalue.  

We define 
\be
H_u^{10} \equiv H_{\bf(1,2,2)}^{({\bf 1,2},\frac{1}{2})}, \,
\overline{\Delta}_u \equiv 
\overline{\Delta}_{\bf(15,2,2)}^{({\bf 1,2},\frac{1}{2})}, \, 
\Delta_u \equiv 
\Delta_{\bf(15,2,2)}^{({\bf 1,2},\frac{1}{2})}, \,
\Phi_u \equiv 
\Phi_{\bf(\overline{10},2,2)}^{({\bf 1,2},\frac{1}{2})}.
\ee
and 
\be
H_d^{10} \equiv H_{\bf(1,2,2)}^{({\bf 1,2},-\frac{1}{2})}, \,
\overline{\Delta}_d \equiv 
\overline{\Delta}_{\bf(15,2,2)}^{({\bf 1,2},-\frac{1}{2})}, \,
\Delta_d \equiv \Delta_{\bf(15,2,2)}^{({\bf 1,2},-\frac{1}{2})}, \, 
\Phi_d \equiv 
\Phi_{\bf(10,2,2)}^{({\bf 1,2},-\frac{1}{2})}.
\ee
In the basis 
$\left\{ H_u^{10}, \overline{\Delta}_u, \Delta_u, \Phi_u \right\}$, 
the mass matrix is written as  
\be
\label{Mdoublet}
M_{\mathsf{doublet}} \equiv 
\left(
\begin{array}{cccc}
\begin{array}{c}
2 m_3 \\
\frac{\lambda_4 \phi_2}{\sqrt{10}} 
-\frac{\lambda_4 \phi_3}{2 \sqrt{5}} \\
-\frac{\lambda_3 \phi_2}{\sqrt{10}} 
-\frac{\lambda_3 \phi_3}{2 \sqrt{5}} \\
\frac{\lambda_3 v_R}{\sqrt{5}}
\end{array}
&\begin{array}{c}
\frac{\lambda_3 \phi_2}{\sqrt{10}} 
-\frac{\lambda_3 \phi_3}{2 \sqrt{5}} \\
m_2 +\frac{\lambda_2 \phi_2}{15 \sqrt{2}} 
-\frac{\lambda_2 \phi_3}{30} \\
0 \\
0
\end{array}
&\begin{array}{c}
-\frac{\lambda_4 \phi_2}{\sqrt{10}} 
-\frac{\lambda_4 \phi_3}{2 \sqrt{5}} \\
0 \\
m_2 +\frac{\lambda_2 \phi_2}{15 \sqrt{2}} 
+\frac{\lambda_2 \phi_3}{30}\\
-\frac{\lambda_2 v_R}{10}
\end{array}
&\begin{array}{c}
\frac{\lambda_4 \overline{v_R}}{\sqrt{5}} \\
0 \\
-\frac{\lambda_2 \overline{v_R}}{10} \\
2 m_1 + \frac{\lambda_1 \phi_2}{\sqrt{2}} 
+ \frac{\lambda_1 \phi_3}{2}
\end{array}
\end{array}
\right).  
\ee
The corresponding mass terms of the superpotential read 
\be
W_m = \left(H_u^{10}, \overline{\Delta}_u, \Delta_u, \Phi_u \right) 
\, M_{\mathsf{doublet}} \,
\left(H_d^{10}, \Delta_d, \overline{\Delta}_d, \Phi_d \right)^{\mathsf{T}}.  
\label{Wm}
\ee
The requirement of the existence of a zero mode leads to the 
following condition.  
\be 
\det{M_{\mathsf{doublet}}} = 0.  
\label{splittings}
\ee
For instance, in case of $\lambda_3 = 0$, 
$m_2 + \frac{\lambda_2 \phi_2}{15 \sqrt{2}} 
- \frac{\lambda_2 \phi_3}{30}=0$, 
we obtain a special solution to Eq. (\ref{splittings}), 
while it keeps a desirable vacuum and it does not produce 
any additional massless fields.  However, we proceed our 
arguments hereafter without using this special solution.  

We can diagonalize the mass matrix, $M_{\mathsf{doublet}}$ 
by a bi-unitary transformation.   
\be
U^{\ast} \,M_{\mathsf{doublet}} \,V^{\dagger}
= {\mathrm{diag}}(0, M_1, M_2, M_3).    
\ee
Then the mass eigenstates are written as  

\bea
\left(H_u, 
\, {\mathsf{h}}_u^1, \, {\mathsf{h}}_u^2, \, {\mathsf{h}}_u^3 \right)
&=& 
\left(H_u^{10}, {\overline{\Delta}}_u, \Delta_u, 
\Phi_u \right) \, U^{\mathsf{T}}, 
\nonumber\\
\left(H_d, 
\, {\mathsf{h}}_d^1, \, {\mathsf{h}}_d^2, \, {\mathsf{h}}_d^3 \right)
&=&
\left(H_d^{10}, \Delta_d, {\overline{\Delta}}_d, 
\Phi_d \right) \, V^{\mathsf{T}}. 
\label{UV}
\eea
Here $H_u,~H_d$ are MSSM light Higgs doublets. 
We get the explicit form of $U$ and $V$ from \bref{Mdoublet}, and thus we can connect the oacillation data with the GUT Yukawa coupling.
Thus the intermediate energy scales are severely constrained from the low energy neutrino data, and the gauge coupling unification 
 at the GUT scale may be spoiled. \\
This fact has been explicitly shown in \cite{Bertolini2}, 
 where the gauge couplings are not unified any more 
 and even the ${\rm SU}(2)$ gauge coupling blows up below the GUT scale. 
Thus the detail analyses of superpotential was the great progress but it reveals the unambiguous detail of structure, which reveals also pathologies.\\
However, this is easily remedied by the addition of ${\bf 120}$ Higgs in Yukawa coupling \cite{Mimura}. We mean that the dominant part may be governed by the minimal SO(10) but
such generalization does not spoil the renormalizable SO(10) GUT yet. 
\paragraph{General Higgs superpotential}
Also we may consider the more general superpotential for completeness \cite{fuku1}. 
\bea
W &=& \frac{1}{2} m_{1} \Phi^2 + m_{2} \overline{\Delta} \Delta + \frac{1}{2} m_{3} H^2 
\nonumber\\
&+& \frac{1}{2} m_{4} A^2 + \frac{1}{2} m_{5} E^2 + \frac{1}{2} m_{6} D^2 
\nonumber\\
&+& \lambda_{1} \Phi^3 + \lambda_{2} \Phi \overline{\Delta} \Delta
+ \left(\lambda_3 \Delta + \lambda_4 \overline{\Delta} \right) H \Phi
\nonumber\\
&+&
\lambda_{5} A^2 \Phi -i \lambda_{6} A \overline{\Delta} \Delta
+ \frac{\lambda_7}{120} \varepsilon A \Phi^2
\nonumber\\
&+& E \left( \lambda_{8} E^2 + \lambda_{9} A^2 + \lambda_{10} \Phi^2 
+ \lambda_{11} \Delta^2 + \lambda_{12} \overline{\Delta}^2 + \lambda_{13} H^2 
\right)
\nonumber\\
&+& D^2 \left( \lambda_{14} E + \lambda_{15} \Phi \right)
\nonumber\\
&+& D \left\{ \lambda_{16} H A + \lambda_{17} H \Phi + \left( 
\lambda_{18} \Delta + \lambda_{19} \overline{\Delta} \right) A 
+ \left( \lambda_{20} \Delta + \lambda_{21} \overline{\Delta} \right) \Phi  
\right\},  
\label{potential}
\eea
Here $A = {\bf 45}$, $\Delta = {\bf 126}$, $\Phi = {\bf 210}$ 
and $E = {\bf 54}$ irreps.  
For general coupling 
constants $\lambda_1,\cdots,\lambda_{21}$, $m_1,\cdots,m_8$, the solutions 
with higher symmetries are specified by following relations.
Solutions with higher symmetries are characterized by:
\begin{enumerate}
\item
$SU(5) \times U(1)_{X}$ and $(SU(5) \times U(1))^{\mathrm{flipped}}$ symmetry solutions
\bea
\label{G51vac}
\left\{
\begin{array}{lll}
E  &=& v_{R} \ =\ 0,
\\
\Phi_1 & = & \frac{\varepsilon}{\sqrt{6}} \,\Phi_3, \quad
\Phi_2 \ =\ \frac{\varepsilon}{\sqrt{2}} \, \Phi_3,
A_{1} \ = \ \frac{2\varepsilon}{\sqrt{6}} A_{2},  
\end{array}
\right. 
\eea
where $\varepsilon = 1$ and $\varepsilon = -1$ correspond to the 
$SU(5)\times U(1)_{X}$ symmetric vacua and 
$(SU(5) \times U(1))^{\mathrm{flipped}}$ symmetric vacua, respectively.
\item
$SU(5)$ symmetry solutions
\bea
\label{G5vac}
\left\{
\begin{array}{lll}
E  &=& 0,
\\
\Phi_1 & = & \frac{1}{\sqrt{6}} \,\Phi_3, \quad
\Phi_2  \ =\ \frac{1}{\sqrt{2}}\, \Phi_3, \quad
A_{1}\ = \ \frac{2}{\sqrt{6}}\, A_{2}, \quad
v_R \ \neq\ 0.
\end{array}
\right.
\eea
\item
$G_{422} \equiv SU(4) \times SU(2)_{L} \times SU(2)_{R}$ symmetry solutions
\bea
\label{G422vac}
\left\{
\begin{array}{lll}
\Phi_2 & =& \Phi_3 \ =\ A_{1}\ =\ A_{2} \ =\ v_{R} \ =\ 0, 
\\
\Phi_1 & \neq& 0, \quad E \ \neq\ 0. 
\end{array}
\right.
\eea
\item
$G_{3221} \equiv SU(3)_{C} \times SU(2)_{L} \times SU(2)_{R} \times U(1)_{B-L}$ symmetry solutions
\bea
\label{G3221vac}
\left\{
\begin{array}{lll}
\Phi_3 &=& A_1 \ =\ v_R \ =\ 0,
\\
\Phi_1  &\neq& 0,  \quad \Phi_2 \ \neq\ 0,  \quad A_2 \ \neq\ 0,  \quad E \ \neq\ 0.
\end{array}
\right. 
\eea
\item
$G_{421} \equiv SU(4) \times SU(2)_{L} \times U(1)$ symmetry solutions
\bea
\label{G421vac}
\left\{
\begin{array}{lll}
\Phi_2 &=& \Phi_3\ =\ A_2\ =\ v_R\ =\ 0 
\\
\Phi_1  &\neq& 0,  \quad A_1 \ \neq\ 0, \quad  E \ \neq\ 0.
\end{array}
\right.
\eea
\item
$G_{3211} \equiv SU(3)_{C} \times SU(2)_{L} \times U(1)_{R} \times U(1)_{B-L}$ 
symmetry solutions
\bea
\label{G3211vac}
\left\{
\begin{array}{lll}
v_{R} &=&0, 
\\
\Phi_i &\neq& 0\ (i=1,2,3), \quad A_i\ \neq\ (i=1,2), \quad E\ \neq\ 0. 
\end{array}
\right. 
\eea
\end{enumerate}
The higher symmetry solutions given in Eqs. (\ref{G51vac})-(\ref{G3211vac}) lead
to the crucial consistency checks for all results in this paper.
In this talk, however, we need the alternative approaches.

In order to avoid this trouble more drastically and keep the successful gauge coupling unification as usual, it is desirable that all Higgs multiplets 
 have masses around the GUT scale, but some Higgs fields 
 develop VEVs at the intermediate scale. 
More Higgs multiplets and some parameter tuning in the Higgs sector 
 are necessary to realize such a situation. 

In addition to the issue of the gauge coupling unification, 
 the minimal SO(10) model potentially suffers from the problem 
 that the gauge coupling blows up around the GUT scale. 
This is because the model includes many Higgs multiplets of 
 higher dimensional representations. 

According to the line of thoughts from \bref{cutoff1} to \bref{cutoff2}, it was natural to consider
\be
L_{GUT}=L_{ren}''+\frac{L_3}{\Lambda_3}.
\label{cutoff3}
\ee
up to $M_{PL}$.
Here $\Lambda_3=O(M_{Pl})$ and gravitation (spacetime structure) appears as a subdominant term. However the blow-up before $M_{PL}$ problem shows that such scheme does not exist in its naive sence. 

 The minimal SO(10) model also is faced on the fast proton decay \cite{detail}

These facts strongly (but not indipensablly) suggest the presence of extra dimensions, which gives not only solve the above problems but also new insights for SUSY breaking mechanism
\cite{Luty}

\section{Part II--SO(10) GUT in 5D}
In this Part we propose a solution to the problem of the minimal SO(10)
discussed in Part I.

\subsection{Minimal SO(10) model in a warped extra dimension} 
 We consider the minimal SUSY SO(10) model 
in the following 5D warped geometry
\footnote{This part is based on the work, 
``Solving problems of 4D minimal SO(10) model in a warped extra dimension'',
T. Fukuyama, T. Kikuchi and N. Okada, Phys.Rev.{\bf D75} 075020 (2007)
[Archive: hep-ph/0702048].}
\begin{eqnarray}
 d s^2 = e^{-2 k r_c |y|} \eta_{\mu \nu} d x^{\mu} d x^{\nu} 
 - r_c^2 d y^2 \; , 
\end{eqnarray}
 for $-\pi\leq y\leq\pi$, where $k$ is the AdS curvature, and 
 $r_c$ and $y$ are the radius and the angle of $S^1$, respectively \cite{RS}. 
The most important feature of the warped extra dimension model 
 is that the mass scale of the IR brane is warped down to 
 a low scale by the warp factor \cite{RS}, $ \omega \equiv e^{-k r_c \pi}$, 
 in four dimensional effective theory. 
For simplicity, we take the cutoff of the original five dimensional theory 
 and the AdS curvature as $M_5 \simeq k \simeq M_{PL}$, 
 the four dimensional Planck mass,
 and so we obtain the effective cutoff scale 
 as $\Lambda_{IR}= \omega M_{PL}$ in effective four dimensional theory. 
Now let us take the warp factor so as for the GUT scale 
 to be the effective cutoff scale 
 $ M_{\rm GUT}= \Lambda_{IR}=\omega M_{PL}$ \cite{Nomura:2006pn}. 
As a result, we can realize, as four dimensional effective theory, 
 the minimal SUSY SO(10) model 
 with the effective cutoff at the GUT scale. 

Before going to a concrete setup of the minimal SO(10) model 
 in the warped extra dimension, 
 let us see Lagrangian for the hypermultiplet in the bulk, 
\begin{eqnarray}
{\cal L} &=& \int dy \left\{ 
\int d^4 \theta \; r_c \; e^{- 2 k r_c |y|} 
 \left( 
 H^{\dagger} e^{- V} H + H^{c} e^{ V}H^{c \dagger} 
 \right) \right. \nonumber \\
&+& 
\left. 
\int d^2 \theta e^{-3 k r_c |y|}
 H^{c} \left[ 
  \partial_{y} - \left( \frac{3}{2}-c \right) k r_c \epsilon(y) 
 - \frac{\chi}{\sqrt{2}}  \right]  
 H  +h.c. \right \} \; , 
\label{bulkL}
\end{eqnarray}
where $c$ is a dimensionless parameter, 
$\epsilon(y)=y/|y|$ is the step function, 
 $H, ~H^c$ is the hypermultiplet charged under some gauge group, 
 and $V, ~\chi$
 are the vector multiplet and the adjoint chiral multiplets, 
 which form an $N=2$ SUSY gauge multiplet. 
 $Z_2$ parity for $H$ and $V$ is assigned as even, 
 while odd for $H^c$ and $\chi$. 

When the gauge symmetry is broken down, 
 it is generally possible that the adjoint chiral multiplet 
 develops its VEV \cite{Kitano:2003cn}. 
Since its $Z_2$ parity is odd, 
 the VEV has to take the form, 
\bea
\left<\Sigma \right> = 2 \alpha k r_c  \epsilon(y) \; , 
\eea
where the VEV has been parameterized by a parameter $\alpha$. 
In this case, the zero mode wave function of $H$ 
 satisfies the following equation of motion:
\bea
\left[\partial_y -
 \left(\frac{3}{2}-c + \alpha \right) k r_c \epsilon(y) \right]H =0 
\eea
which yields 
\bea
H = \frac{1}{\sqrt{N}} 
 e^{ (3/2-c + \alpha) kr_c |y|} \; h(x^\mu) \; , 
\eea
where $h(x^\mu)$ is the chiral multiplet in four dimensions. 

Lagrangian for a chiral multiplets on the IR brane is given by 
\bea 
 {\cal L}_{IR}=  
 \int d^4 \theta \;  \omega^\dagger \omega \;  \Phi^\dagger \Phi 
 +\left[   \int d^2 \theta \;  \omega^3 \; W(\Phi) + h.c.   
 \right] \; ,
\eea 
where we have omitted the gauge interaction part 
 for simplicity. 
If it is allowed by the gauge invariance, 
 we can write the interaction term 
 between fields in the bulk and on the IR brane, 
\bea 
{\cal L}_{int}= \int d^2 \theta \omega^3 
 \frac{Y}{\sqrt{M_5}} \Phi^2 H(y=\pi) +h.c. \; ,  
\label{IR-Yukawa} 
\eea  
where $Y$ is a Yukawa coupling constant, 
 and $M_5$ is the five dimensional Planck mass 
 (we take $M_5\sim M_{PL}$ as mentioned above, for simplicity). 
Rescaling the brane field $\Phi \rightarrow  \Phi/\omega$ 
 to get the canonically normalized kinetic term 
 and substituting the zero-mode wave function of the bulk fields, 
 we obtain Yukawa coupling constant 
 in effective four dimensional theory as 
\bea 
  Y_{4D} \sim Y 
\eea 
 if $e^{ (1/2 - c +  \alpha ) k r_c \pi}  \gg 1$,  
 while 
\bea 
  Y_{4D} \sim Y 
  \times e^{ (1/2 -  c +  \alpha ) k r_c \pi}  \ll Y \; , 
\label{suppression}
\eea  
 for $e^{ (1/2 - c + \alpha ) k r_c \pi }  \ll 1$. 
In the latter case, we obtain a suppression factor 
 since $H$ is localized around the UV brane. 

Now we give a simple setup of the minimal SO(10) model 
 in the warped extra dimension. 
We put all ${\bf 16}$ matter multiplets on the IR ($y=\pi$) brane, 
 while the Higgs multiplets ${\bf 10}$ and $\overline{\bf 126}$ 
 are assumed to live in the bulk. 
In Eq.~(\ref{IR-Yukawa}), replacing the brane field into the matter 
 multiplets and the bulk field into the Higgs multiplets, 
 we obtain Yukawa couplings in the minimal SO(10) model. 
The Lagrangian for the bulk Higgs multiplets are given 
 in the same form as Eq.~(\ref{bulkL}), 
 where $\chi$ is the SO(10) adjoint chiral multiplet, ${\bf 45}$. 
As discussed above, since the SO(10) gauge group is broken 
 down to the SM one, 
 some components in $\chi$ which is singlet under the SM gauge group 
 can in general develop VEVs. 
Here we consider a possibility that 
 the ${\rm U}(1)_X$ component 
 in the adjoint $\chi ={\bf 45}$ under the decomposition 
 SO(10) $\supset {\rm SU}(5) \times {\rm U}(1)_X$ has 
 a non-zero VEV,
\bea
 {\bf 45} = {\bf 1}_0
 \oplus {\bf 10}_{+4} \oplus \overline{\bf 10}_{-4}
 \oplus {\bf 24}_0 \; .   \nonumber 
\eea
The $\overline{\bf 126}$ Higgs multiplet 
 are decomposed under ${\rm SU}(5) \times {\rm U}(1)_X$ as
\bea
\overline{\bf 126} &=& 
 {\bf 1}_{+10} 
 \oplus {\bf 5}_{+2} \oplus \overline{\bf 10}_{+6}
 \oplus {\bf 15}_{-6} 
 \oplus \overline{\bf 45}_{-2} \oplus {\bf 50}_{+2} \; . 
 \nonumber  
\eea
In this decomposition, 
 the coupling between a bulk Higgs multiplet and 
 the ${\rm U}(1)_X$ component in $\chi$ is proportional 
 to  ${\rm U}(1)_X$ charge, 
\be
{\cal L}_{int} \supset \frac{1}{2} \int d^2 \theta \omega^3
Q_X \langle \Sigma_X \rangle H^c H + h.c. \;,
\ee
 and thus each component effectively obtains 
 the different bulk mass term,  
\bea 
  \left( \frac{3}{2} - c \right) k r_c  
   + \frac{1}{2}Q_X \langle \Sigma_X \rangle,   
\eea
 where $Q_X$ is the ${\rm U}(1)_X$ charge of corresponding Higgs multiplet, 
 and $\Sigma_X$ is the scalar component of 
 the ${\rm U}(1)_X$ gauge multiplet (${\bf 1}_0$). 
Now we obtain different configurations of the wave functions 
 for these Higgs multiplets. 
Since the ${\bf 1}_{+10}$ Higgs has a large ${\rm U}(1)_X$ charge 
 relative to other Higgs multiplets, 
 we can choose parameters $c$ and $\langle \Sigma_X \rangle$ 
 so that Higgs doublets are mostly localized around the IR brane 
 while the ${\bf 1}_{+10}$ Higgs is localized around the UV brane. 
Therefore, 
 we obtain a suppression factor 
 as in Eq.~(\ref{suppression}) 
 for the effective Yukawa coupling between 
 the Higgs and right-handed neutrinos. 
In effective four dimensional description, 
 the GUT mass matrix relation is partly broken down, 
 and the last term in Eq.~(\ref{Yukawa3}) is replaced into 
\bea 
 Y_{126}^{ij} v_R \rightarrow Y_{126}^{ij} (\epsilon v_R) \; ,  
\eea
where $\epsilon$ denotes the suppression factor. 
By choosing an appropriate parameters 
 so as to give $\epsilon=10^{-2}-10^{-3}$, 
 we can take $v_R \sim M_{\rm GUT}$ 
 and keep the successful gauge coupling unification in the MSSM.

Thus, in order to solve the problems in 4D, we have considered 
 the minimal SO(10) model in the warped extra dimension. 
As a simple setup, we have assumed that matter multiplets 
 reside on the IR brane 
 while the Higgs multiplets reside in the bulk. 
The warped geometry leads to a low scale effective cutoff
 in effective four dimensional theory, 
 and we fix it at the GUT scale. 
Therefore, the four dimensional minimal SO(10) model 
 is realized as the effective theory with the GUT scale cutoff.

However, it gives rise another problem:
SO(10) is anomaly free theory and there appears no D term as far as we consider spontaneously broken scenario. So we can not cancell D term caused by the vev of ${\bf 45}$ either at bulk and branes. We did not propose the mechanism how ${\bf 45}$ has vev. Also it was not clear about the dangerous proton decay. Of course we overlooked some point and we do not exclude this scenario.

In the next section we consider another possibility, orbifold GUT model.
We break the original
N=2 (in the sence of 4D) SUSY SO(10) invariant theory into $SU(4)_C\times SU(2)_L\times SU(2)_R$ (hereafter PS for short) not spontaneously but by the boundary conditions in the orbifold $S^1/{Z_2\times Z_2'}$.
If we consider extra dimensions, the chiral fields need orbifold like compactification \cite{JEKim}. 
There are so many papers in this region. We consider the most simple and clearcut scenario in this paper.

\subsection{Orbifold GUT} 

We consider a SUSY $SO(10)$ SUSY GUT in 5D orbifold. \footnote{This part is based on the work by
T. Fukuyama and N. Okada,"A simple SO(10) GUT in five dimensions" [arXive:hep-ph/0803.1758].} 
Usually SO(10) is considered in six dimension, whereas SU(5) in 5 dimension.
This is because we need at least two projections for SO(10) down to SM \cite{JEKim}\cite{6D} if we break the symmetry only through boundary conditions.
It should be remarked that even in this case we need the Higgs mechanism to break SM to $SU(3)_c\times U(1)_{em}$. So dimensionality 6 does not have definite meaning.

It is very important that in the PS brane we can discard $({\bf 6},{\bf 2},{\bf 
1})$ as mentioned in the previous section, and harmful proton decay is circumvented without boundary condition.
As mentioned, orbifold is essential for chiral dynamics in extra dimension.
However, 
more extra dimensions are not indispensable in GUT framework. Our set up also indicates 
why the PS brane is visible brane, where the PS (and not SO(10)) is broken by Higgs 
mechanism. 

In 4D language, 5D vector multiplet consists of N=1 vector supermultiplet
V and an N=1 chiral multiplet $\Psi$. In SO(10) $V$ is ${\bf 45}$.

One extra dimension is compactified on the orbifold $S^1/{Z_2\times Z_2'}$ \cite{kawamura}. 
That is, N=2 SO(10) invariant action in 5D is decomposed into
N=1 SO(10) invariant $y=0$ brane and the PS invariant brane at $y=\pi R/2$
by the boundary conditions of bulk gauge (See Table I).
We do consider neither matter nor Higgs in the bulk. For the different set up, see \cite{Hall} \cite{Raby} \cite{Barr}.

\begin{table}[h]
\begin{tabular}{|c|c|c|}
\hline
& & \\
$(P,P')$ & field & mass\\
& & \\
\hline
& & \\
$(+,+)$ &  $V(15,1,1)$, V(1,3,1),V(1,1,3) & $\frac{2n}{R}$\\
& & \\
\hline
& & \\
$(+,-)$ &  $V(6,2,2)$ & $\frac{(2n+1)}{R}$ \\
& & \\
\hline
& & \\
$(-,+)$ &  $\Phi (6,2,2)$
& $\frac{(2n+1)}{R}$\\
& & \\
\hline
& & \\
$(-,-)$ &  $\Phi (15,1,1)$, $\Phi (1,3,1)$, $\Phi (1,1,3)$ & $\frac{(2n+2)}{R}$ \\
& & \\
\hline
\end{tabular}
\caption{P and P' assignment and masses ($n\ge 0$) of fields in the vector multipet $(V,~\Phi)$ under the PS group. P' even $V$ contains the PS gauge
bosons.}
\label{t1}
\end{table}
\begin{table}[h]
\begin{tabular}{|c|c|}
\hline
& brane at $y=\pi R/2$ \\ 
\hline
& \\
Matter Multiplets & $\psi_i=F_{Li} \oplus F_{Ri} \quad (i=1,2,3)$ \\
 & \\
\hline
 & \\
Higgs Multiplets & 
$({\bf 1},{\bf 2},{\bf 2})_H$,  
$({\bf 1},{\bf 2},{\bf 2})'_H$,
$({\bf 15},{\bf 1},{\bf 1})_H$,
$({\bf 6},{\bf 1},{\bf 1})_H$ \\  & 
$({\bf 4},{\bf 1},{\bf 2})_H$, 
$(\overline{{\bf 4}},{\bf 1},{\bf 2})_H$, 
$({\bf 4},{\bf 2},{\bf 1})_H$, 
$(\overline{{\bf 4}},{\bf 2},{\bf 1})_H$  \\
& \\
\hline
\end{tabular}
\caption{
Particle contents on the PS brane. 
Here, we impose the left-right symmetry. 
}
\label{localization}
\end{table}

The PS invriance in the PS brane is broken by the Higgs mechanism down to the Standard model by the PS subgroup of $({\bf 4},{\bf 1},{\bf 2})_H$ ($H_R$ terms in \bref{Yukawa}).
In the SO(10) model\cite{M-N-F}, the left- (right-) handed fermions 
$\psi_{L(R)i}$ in a given i-th generation are assigned to a single irreducible 
{\bf 16}.
Since ${\bf 16}\times {\bf 16} = {\bf 10}_S +{\bf 120}_A + 
{\bf 126}_S$,  the fermion  masses are 
generated when the Higgs fields of ${\bf 10}$, and ${\bf 120}$, and 
$\overline{{\bf 126}}$ dimensional SO(10) representation (denoted by 
$\phi_{10}$, $\phi_{120}$, and $\phi_{126}$, respectively) develop 
nonvanishing expectation values. Their decomposition under 
$SU(4)\times SU_L(2)\times SU_R(2)$ are given by 
\begin{eqnarray}
{\bf 10}  &=& ({\bf 6},{\bf 1},{\bf 1}) +
	({\bf 1},{\bf 2},{\bf 2}), \nonumber \\
{\bf 120} &=& ({\bf 15},{\bf 2},{\bf 2})+({\bf 6},{\bf 3},{\bf 1})+
	(\overline{{\bf 6}},{\bf 1},{\bf 3})+({\bf 1},{\bf 2},{\bf 2})+
		({\bf 20},{\bf 1},{\bf 1}), \\
{\bf 126} &=& ({\bf 10},{\bf 3},{\bf 1})+
	(\overline{{\bf 10}},{\bf 1},{\bf 3}) +
		({\bf 15},{\bf 2},{\bf 2})+
		(\overline{{\bf 6}},{\bf 1},{\bf 1}). \nonumber
\end{eqnarray}
On the other hand, the fermion field of 16-dimensinal SO(10) representation is 
decomposed as 
\begin{eqnarray}
{\bf 16} & =({\bf 4},{\bf 2},{\bf 1})+(\overline{{\bf 4}},{\bf 1},{\bf 2}).
\end{eqnarray}
With respect to $SU(4)\times SU_L(2)\times SU_R(2)$, the left - and right- handed  
quarks and leptons of a given i-th generation are assigned as
\begin{eqnarray}
	\left(
		\begin{array}{cccc}
		u_r & u_y & u_b & \nu_e \\
		d_r & d_y & d_b & e
		\end{array}
	\right)_{L(R)}	\equiv F_{L(R)1} ,
\end{eqnarray}
\(F_{L(R)2}\) and \(F_{L(R)3}\) are likewise defined for 
the 2nd and 3rd generations.
Note that their transformation properties are
$F_{Li}=({\bf 4},{\bf 2},{\bf 1})$ 
and $F_{Ri}=({\bf 4},{\bf 1},{\bf 2})$ 
and that ($F_{Li} + \overline{F_{Ri}}$ ) 
yields the {\bf 16} of SO(10).
Since $({\bf 4},{\bf 2},{\bf 1})\times
(\overline{{\bf 4}},{\bf 1},{\bf 2})=
({\bf 15},{\bf 2},{\bf 2})+({\bf 1},{\bf 2},{\bf 2})$, 
the Dirac masses for quarks and leptons are generated by $({\bf 15},{\bf 2},{\bf 2})_H+({\bf 1},{\bf 2},{\bf 2})_H$.

From $\tau-b$ unification at GUT scale the third generation is described by $({\bf 1},{\bf 2},{\bf 2})$ of $H_{10}$. The deviations of the first and second generations are complimented by the
$({\bf 15},{\bf 2},{\bf 2})_H~(\mbox{here it is constructed from}~({\bf 15},{\bf 1},{\bf 1})\times ({\bf 1},{\bf 2},{\bf 2})': \mbox{see Table 4})$.
and
\bea
({\bf 15},{\bf 2},{\bf 2})&=&({\bf 1},{\bf 2},{\bf 1/2})+({\bf 1},{\bf 2},-{\bf 1/2})+({\bf 8},{\bf 2},{\bf 1/2})+({\bf 8},{\bf 2},-{\bf 1/2})+({\bf 3},{\bf 2},{\bf 1/6})+(\overline{{\bf 3}},{\bf 2},-{\bf 1/6})\nonumber\\
&+&({\bf 3},{\bf 2},{\bf 7/6})+(\overline{{\bf 3}},{\bf 2},-{\bf 7/6})
\label{15}
\eea

It is remarkable that the component $({\bf 6},{\bf 1},{\bf 1})$ which was harmful for SO(10) invariant Yukawa coupling does not appear \cite{Mohapatra}, and therefore dimension five operator too.

The spectrum of the PS phases were fully discussed in \cite{Fukuyama:2004xs}. 
The third, fourth octet and the last two triplets in \bref{15} becomes massive by the coupling with the counter parts of another $({\bf 15},{\bf 2},{\bf 2})$.
The fifth and sixth triplet become NG bosons.

SU(4) adjoint 15 have a basis, diag$(1,1,1,-3)$                                 so as to satisfy the traceless condition.  Putting leptons into the 4th color, we get,  so called, eGeorgi-Jarslkogf factor,  $-3$         for leptons. 
Unlike the case of SO(10), the mass matrix forms of $M_L$ and $M_R$ belong to groups different to each other and different from charged fermions.
We may write it in the more familiar forms
\begin{eqnarray}
 M_u &=& c_{10} M_{1,2,2}+ c_{15} M_{15,2,2} \; , 
 \nonumber \\
 M_d &=& M_{1,2,2} + M_{15,2,2} \; ,   
 \nonumber \\
 M_D &=& c_{10} M_{1,2,2} - 3 c_{15} M_{15,2,2} \; , 
 \nonumber \\
 M_e &=& M_{1,2,2} - 3 M_{15.2,2} \; , 
 \nonumber \\
 M_L &=& c_L M_{10,3,1} \; ,
 \nonumber \\ 
 M_R &=& c_R M_{10,1,3} \;, 
\label{massmatrix2}
\end{eqnarray} 

Here the effective $({\bf 10},{\bf 1},{\bf 3})$ was given by $H_RH_R/M_5$ as \bref{Yukawa}.
Otherwise renormalization group equation (RGE) does not converge as we will show in the next paragraph.
Thus in contrast with renormalizable SO(10) where $\overline{{\bf 126}}$ takes part in all mass matrices (See \bref{massmatrix}), there appear independent $M_{15,2,2},~M_{10,1,3},~M_{10,3,1}$ in \bref{massmatrix2} and low energy data fitting is more easily satisfied. 

In the following conveniences, let us introduce the following notations:
\bea
H_1&=&({\bf 1},{\bf 2},{\bf 2})_H, ~H_1^{\prime}=({\bf 1},{\bf 2},{\bf 2})'_H,
\nonumber \\
H_6&=&({\bf 6},{\bf 1},{\bf 1})_H, ~H_{15}=({\bf 15},{\bf 1},{\bf 1})_H,
\nonumber \\ 
H_L&=&({\bf 4},{\bf 2},{\bf 1})_H,
~\overline{H_L} =(\overline{{\bf 4}},{\bf 2},{\bf 1})_H,  
\nonumber \\
H_R&=&({\bf 4},{\bf 1},{\bf 2})_H,
~\overline{H_R}=(\overline{{\bf 4}},{\bf 1},{\bf 2})_H.
\eea

Superpotential relevant for fermion masses is given by%
\footnote{
For simplicity, we have introduced only minimal terms 
 necessary for reproducing observed fermion mass matrices. 
}
\bea
W_Y&=&Y_{1}^{ij}F_{Li}\overline{F_{Rj}}H_1
+\frac{Y_{15}^{ij}}{M_5} F_{Li} \overline{F_{Rj}}
 \left(H_1^{\prime} H_{15} \right) \nonumber\\ 
&+&\frac{Y_R^{ij}}{M_5}\overline{F_{Ri}}
 \overline{F_{Rj}} \left(H_R H_R \right) 
 +\frac{Y_L^{ij}}{M_5} F_{Li}F_{Lj} 
 \left(\overline{H_L} \overline{H_L} \right), 
\label{Yukawa}
\eea 

On the other hand, the $(\overline{{\bf 10}},{\bf 3},{\bf 1})$ and 
$({\bf 10},{\bf 1},{\bf 3})$ in $\phi_{126}$ were responsible 
for the left- and the right- handed Majorana neutrino masses 
and the sama $\phi_{126}$ commited in the Yukawa coupling.
This gave the severe constraints in the minimal SO(10) GUT.
However in the present case $Y_R^{ij}$ are coming from $\overline{H_R}\overline{H_R}$ and are quite independen on the other Dirac Yukawa couplings and we hve no problem in low energy data fitting including the neutrino oscillation data. 
Hereafter we consider type I seesaw since it is sufficient for the neutrino data fitting as indicated just above, and neglect the last term in \bref{Yukawa}.
We introduce Higgs superpotential invariant under the PS symmetry 
 such as 
\bea
W &=& 
 \frac{m_1}{2} H_1^2 + \frac{m_1^\prime}{2} H_1^{\prime 2} 
 + m_{15}~{\rm tr}\left[H_{15}^2 \right] 
  +m_4 \left(\overline{H_L}H_L+\overline{H_R}H_R\right) \nonumber\\
&+& 
\left(H_L \overline{H_R}+ \overline{H_L} H_R \right) 
\left( \lambda_1 H_1 + \lambda_1^\prime H_1^\prime \right) 
+\lambda_{15} \left(\overline{H_R} H_R + \overline{H_L} H_L\right) 
H_{15} \nonumber\\
&+&
\lambda~{\rm tr}\left[H_{15}^3 \right]+
\lambda_6 
 \left( H_L^2+ \overline{H_L}^2 + H_R^2 + \overline{H_R}^2 \right) 
 H_6 .
\label{HiggsW}
\eea
Parameterizing 
 $ \langle H_{15} \rangle =\frac{v_{15}}{2 \sqrt{6}} 
 {\rm diag}(1,1,1,-3)$, 
 SUSY vacuum conditions from Eq.~(\ref{HiggsW}) and 
 the D-terms are satisfied by solutions,  
\bea
v_{15} =\frac{2 \sqrt{6}}{3 \lambda_{15}} m_4,~~~
\langle           H_R  \rangle = 
\langle \overline{H_R} \rangle = 
\sqrt{
 \frac{8 m_4}{3 \lambda_{15}^2} 
 \left( m_{15} -\frac{\lambda}{\lambda_{15}} m_4 \right) }
 \equiv v_{PS} 
\eea 
and others are zero, by which the PS gauge symmetry is broken 
 down to the SM gauge symmetry.  
We choose the parameters so as to be 
 $ v_{15} \simeq \langle H_R \rangle = \langle \overline{H_R} \rangle$.  
Note that the last term in Eq.~(\ref{HiggsW}) is necessary 
 to make all color triplets in $H_R$ and $\overline{H_R}$ heavy.

Weak Higgs doublet mass matrix is given by
\begin{equation}
\left(
       \begin{array}{ccc}
        H_1, & H_1^\prime, & H_L \end{array}
\right)\left(
        \begin{array}{ccc}
        m_1 &  0         & \lambda_1        \langle H_R \rangle \\
        0   & m_1^\prime & \lambda_1^\prime \langle H_R \rangle \\ 
        \lambda_1 \langle \overline{H_R} \rangle &  
        \lambda_1^\prime \langle \overline{H_R} \rangle & m_4 
        \end{array}
\right)\left(
        \begin{array}{c}
        H_1\\
        H_1^\prime \\
        \overline{H_L}
        \end{array}
\right).
\end{equation} 
In order to realize the MSSM at low energy, 
 only one pair of Higgs doublets out of the above tree pairs 
 should be light, while others have mass of the PS symmetry breaking scale. 
This doublet-doublet Higgs mass splitting requires 
 the fine tuning of parameters to satisfy 
\bea
\det M=m_1 m_1^\prime m_4 - 
 (m_1 \lambda_1^{\prime 2} + m_1^\prime \lambda_1^2) v_{PS}^2=0.
\label{vev}
\eea

\vspace{1cm}

\paragraph{RGE}
In our set up, the evolution of gauge coupling has three stages, $G_{321}$ (SM+MSSM), $G_{422}$ (the PS) and $M_c~(\equiv (1/R))$ stages.
In the $G_{321}$ stage, the beta functions $b_i$ are defined by

\be
\frac{1}{\alpha_i (\mu)}=\frac{1}{\alpha_i(M)}+\frac{1}{2\pi}b_i\mbox{ln}\left(
\frac{M}{\mu}\right).~~(i=3,2.1)
\ee
$b_i$ at $G_{321}$ are
\be
b_3=-7,~b_2=-19/6,~b_1=41/10
\ee
at $M_{SUSY}>\mu>M=M_Z$ and
\be
b_3=-3,~b_2=1,~b_1=33/5
\ee
at $M_c>\mu>M=M_{SUSY}$
The PS symmetry is recovered at $\mu=v_{PS}$. However, we assume $v_{PS}=M_c$ for simplicity, and the matching condition holds
\bea
\alpha_3^{-1}(M_c)&=&\alpha_4^{-1}(M_c)\nonumber\\
\alpha_2^{-1}(M_c)&=&\alpha_{2L}^{-1}(M_c)\\
\alpha_1^{-1}(M_c)&=&[2\alpha_4^{-1}(M_c)+3\alpha_{2R}^{-1}]/5\nonumber
\label{maching}
\eea
at $M_c$. In the PS stage $\mu>M_c$, the threshold corrections $\Delta_i$ due to KK mode in the bulk are added,
\be
\frac{1}{\alpha_i (\mu)}=\frac{1}{\alpha_i(M_c)}+\frac{1}{2\pi}b_i\mbox{ln}\left(
\frac{M_c}{\mu}\right)+\Delta_i.~~(i=4,2_L,2_R)
\ee
The beta functions of the PS gauge coupling constants are from the contents of Tables 3 and 4
\be
b_4=3, ~b_{2L}=b_{2R}=6.
\label{lrs}
\ee


$\Delta_i$ are
\be
\Delta_i=\frac{1}{2\pi}b_i^{even}\sum_{n=0}^{N_l}\theta(\mu-(2n+2)M_c)\mbox{ln}\frac{(2n+2)M_c}{\mu}
+\frac{1}{2\pi}b_i^{odd}\sum_{n=0}^{N_l}\theta(\mu-(2n+1)M_c)\mbox{ln}\frac{(2n+1)M_c}{\mu}
\ee
with 
\bea
b_i^{even}&=&(-8,-4,-4)\nonumber\\
b_i^{odd}&=&(-8,-12,-12)
\eea
under $G_{422}$.

In Fig.3 we depict the gauge coupling unification for left-right symmetric case.
For simplicity we first assumed $M_c=v_{PS}$.
The point $v_{PS}$ is given by $\alpha_{2L}^{-1}(v_{PS})=\alpha_2^{-1}(v_{PS})=\alpha_{2R}^{-1}(v_{PS})=(5\alpha_1^{-1}-2\alpha_3)/3|_{v_{PS}}$.
$M_*$ is the scale of gauge coupling unification.
The result is 
\be
v_{PS}=M_c=1.19\times 10^{16} \mbox{GeV},~~M_{GUT}=4.61\times 10^{17} \mbox{GeV}\ee

In SO(10) model in 4D, $v_{PS}=O(10^{14}\mbox{GeV})$ was preferable for neutrino masses. However, it is not the case for the theories of SO(10) in 5D or the PS model in 4D.
As is easily seen from \bref{Yukawa},
\be
M_R\sim Y_Rv_{PS}^2/M_5\sim 0.1Y_Rv_{PS}
\ee
So if we assume $Y_R=0.1$ we obtain the reasonable value of light neutrino mass.
\begin{figure}[h]
{\includegraphics*[width=.6\linewidth]{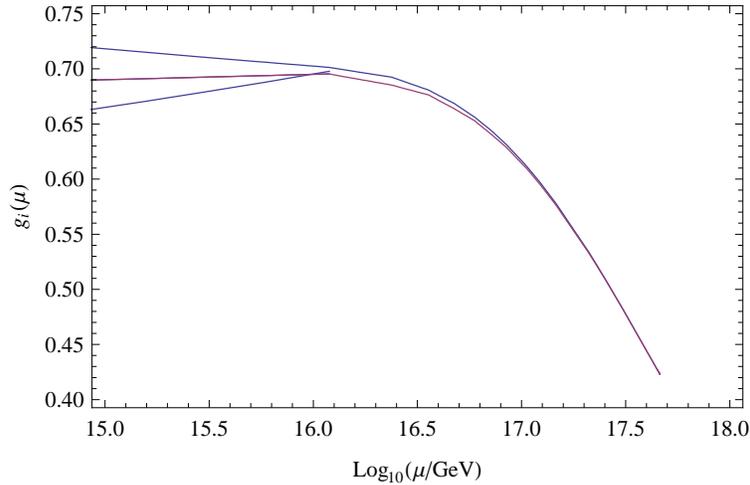}
\label{Fig1}}
\caption{
Gauge coupling unification in the left-right symmetric case. 
Each line from top to bottom corresponds to 
 $g_3$, $g_2$ and $g_1$ for $ \mu < M_c$,
 while  $g_3=g_4$ and $g_2=g_{2R}$ for $ \mu > M_c$. 
}
\end{figure}

For the case of no left-right symmetry,
we can not fix $v_{PS}$ as before. $v_{PS}$ and $M_c$ are determined from the reqirement of gauge coupling unification.

\paragraph{gaugino mediation and gravitino problem}
At $y=0$ brane, SO(10) remaind invariant but supersymmetry is broken by the F-term of SO(10) singlet $S,~F_S\theta^2$.
The interaction between $S$ and bulk fields is
\bea
L&=&\delta (y)\int d^2\theta\lambda\frac{S}{M_5^2}\mbox{tr}[W_aW_a], 
\label{mu}
\eea
where $\lambda$ is a dimensionless coupling constant.
The bulk gaugino first obtains the mass
\be
M_{\lambda}=\frac{\lambda F_SM_c}{M_5^2}\approx\frac{\lambda F_SM_5}{M_{PL}^2},
\ee
where we have used $M_5^3/M_c\approx M_{PL}^2$ in the last equality, and $M_c$ comes from the wave function normalization of the bulk gaugino. 

In our scenario, $M_5$ is roughly one order of magnitude 
 smaller than the (reduced) Planck mass ($M_{PL}$). 
As usual, we take $M_{\lambda}=100$GeV$-1$TeV. With this gaugino mass at high energy scale, SUSY breaking mass terms of sfermions are automatically generated through its RGE running and flabour blind cite{Giudice}. Comparing the gaugino mass to gravitino mass $m_{3/2}\approx F_S/M_{PL}$, a typical gaugino mass is smaller than the gravitino mass by a factor $\lambda M_5/M_{PL}\approx 0.1\lambda.$
However, in this simple setup, it turns out that stau 
 is the lightest superpartner (LSP), 
 which is problematic in cosmology \cite{Kawasaki}. 
It has been found that when $M_c > M_{GUT}$, 
 the RGE running in a unified theory pushes up 
 stau mass and leads neutralino to be the LSP \cite{gMSB2}. 
However, in our model, we cannot take such an arrangement, 
 because $M_c$ and $M_{GUT}$ are fixed as $M_c < M_{GUT}$ 
 to realize the gauge coupling unification. 
In order to avoid this problem, 
 we need to extend the SUSY breaking sector. 
It is possible to introduce the gauge mediation \cite{GMSB} 
 on the PS brane, in which gravitino is normally the LSP. 
In general, we can introduce the messenger sector 
 on the brane at $y=0$. 
This setup is basically the same as in Ref.~\cite{MP}, 
 where the gauge mediation was calculated in 5D 
 with the messenger sector on one brane, 
 sfermions on the other brane 
 and gauge multiplets in the bulk. 
When the messenger scale is larger than the compactification scale
 ($M_{mess} > M_c$), 
 the gaugino mass is given by the same formula as in 4D, 
\bea 
 M_{\lambda} \simeq \frac{\alpha_{GUT}}{4 \pi} \frac{F_S}{M_5}, 
\eea 
while sfermion masses are roughly given by 
\bea 
 \tilde{m}^2 \simeq M_{\lambda}^2 \left(\frac{M_c}{M_5} \right)^2.  
\eea 
The sfermion mass squared is suppressed relative to 
 the gaugino mass by a geometric factor $M_c/M_{mess}$, 
 at the messenger scale. 
At low energy, sfermion masses comparable 
 to the gaugino mass are generated through the RGE running. 
In this setup, we find 
\bea 
 \frac{m_{3/2}}{M_{\lambda}} \simeq 
 \frac{M_{mess}}{\frac{\alpha_{GUT}}{4 \pi}M_{PL}} 
 > 10 
\eea 
for $M_{mess} \geq M_c$. 
Thus, in oder to have gravitino the LSP, 
 the messenger scale should be smaller 
 than the compactification scale. 
In this case, soft mass formulas are reduced into 
 the usual four dimensional ones 
 in the gauge mediation scenario. 
 
\vspace{1cm}
\subsection{Conclusion}
We have reviewed the present staus of renormalizable minimal SO(10) GUT
and tried to solve the problems by extending the theory in 4D into 5D.
In the case of warped GUT, it not only solves the problems but also cures the blowup problem of the unified coupling after GUT,
which had been thought as the fatal deficit of renormalizable SO(10) GUT from the side of perturbative SO(10) GUT group.
Other problematic point of high dimension Higgs, if any, is that the complexity of intermediate energy scales (${\bf 210}$ has three SM singlets (22)-(24))
is transformed to the variety of Higgs profile in warped 5D, or is relaxed by using the freedom of the PS invariance in the case of orbifold GUT and gauge coupling unification is recovered.
SO(10) GUT in 5D may also give the device of SUSY breaking mechanism, which was given by hand in 4D.
Probably the more elaborate theory may be the warped orbifold GUT.
The final theory may lead to 10D superstring or heterotic string theories \cite{JEKim2}\cite{Kobayashi}\cite{JEKim} but 5D GUT may give the essential picture towards it.
We should offer more elaborate arguments why an etradimension is needed.
Maldacena conjecture \cite{Maldacena} is very suggestive for it.

Lastly I want to emphasize that the renormalizable minimal SO(10) GUT in 4D still remains valid as the essential part of the future complete theory. It is indeed still premature but "Don't throw the baby out of with the bath water."

\begin{theacknowledgments}
  As the chair of organizers, the author is grateful to Yukawa Institute for Theoretical Physics, Kyoto University and Ritsumeikan University for their finantial supports.
He also thanks all the organizers and participants at this Workshop.
\end{theacknowledgments}


\begin{thebibliography}{0}
\bibitem{Murayama}
H. Murayama, How the neutrino physics requires the theory beyond SM is comprehensively discussed in this Proceedings.
\end{thebibliography}
\end{document}